\documentclass[journal, onecolumn, draftcls,12pt]{IEEEtran}
\usepackage{amssymb,amsmath,epsfig,graphicx,theorem}
\usepackage{amsfonts}
\usepackage{bm}
\usepackage{arydshln}
\usepackage[normalem]{ulem}
\usepackage{algorithm}
\usepackage{algorithmic}
\newtheorem{Theorem}{Theorem}
\newtheorem{Lemma}{Lemma}

\newtheorem{Definition}{Definition}
\newtheorem{Remark}{Remark}

\usepackage{epsfig,rotating,setspace,latexsym,amsmath,epsf,amssymb,bm,color}
\usepackage{cite}

\IEEEoverridecommandlockouts

\title{The Capacity Region of Distributed Multi-User Secret Sharing under The Perfect Privacy Condition}

\author{Jiahong Wu, Nan Liu, Wei Kang%
\thanks{J. Wu and N. Liu are with the National Mobile Communications Research Laboratory,
Southeast University, Nanjing, China (email: \{jiahongwu,nanliu\}@seu.edu.cn). W. Kang is with the School of Information Science and Engineering,
Southeast University, Nanjing, China (email: wkang@seu.edu.cn). }%
\thanks{This work was partially supported by the National Natural Science Foundation of China
under Grants $61971135$, the Research Fund of National Mobile Communications Research Laboratory, Southeast University (No. 2020A03) and Six talent peaks project in Jiangsu Province.
}
}

\begin{document}

\maketitle
\begin{abstract}
We study the distributed multi-user secret sharing (DMUSS) problem under the perfect privacy condition. In a DMUSS problem, multiple secret messages are deployed and the shares are offloaded to the storage nodes. Moreover, the access structure is extremely incomplete, as the decoding collection of each secret message has only one set, and by the perfect privacy condition such collection is also the colluding collection of all other secret messages. The secret message rate is defined as the size of the secret message normalized by the size of a share. We characterize the capacity region of the DMUSS problem when given an access structure, defined as the set of all achievable rate tuples.

In the achievable scheme, we assume all shares are mutually independent and then design the decoding function based on the fact that the decoding collection of each secret message has only one set. Then it turns out that the perfect privacy condition is equivalent to the full rank property of some matrices consisting of different indeterminates and zeros. Such a solution does exist if the field size is bigger than the number of secret messages. Finally with a matching converse saying that the size of the secret is upper bounded by the sum of sizes of non-colluding shares, we characterize the capacity region of DMUSS problem under the perfect privacy condition.
\end{abstract}


\section{introduction}
In classical information-theoretic cryptography with a scenario that we need to send a message to a legal user in the presence of wiretappers, we can encrypt the message with a random key by linear combination and send the ciphertext to the legal user. As the ciphertext and the message are statistically independent, a wiretapper who eavesdrops the ciphertext knows nothing about the message, which is the privacy condition. On the other hand, a legal user can recover the  message with enough information about the random key and the ciphertext still by linear combination, which is the correctness condition. 

Such model \cite{Shannon1949CommunicationTO} was generalized to \emph{secret sharing} by Blakley \cite{Blakley1979SafeguardingCK} and Shamir \cite{shamir1979how}. A secret sharing scheme is a method to share a secret among a set of  users such that only certain subsets of users can recover the secret while any other subset obtains nothing, the collection of the former subsets is called the decoding collection and the collection of the later subsets is referred as the colluding collection. 

Most cryptographic protocols involving secret sharing assume that the central user, called the dealer, has a direct reliable and secure communication channel to all users. In this way, it is assumed that the shares of secret, once computed by the dealer, are readily available to the users. However in a distributed storage scenario, the dealer can be treated as a master node controlling a set of storage nodes and each user has access to a certain subset of storage nodes. Then consider a multi-user secret sharing scenario, in the sense that there is a designated secret message for each user. The user needs to recover the corresponding secret message via accessing the shares in the storage nodes successfully, which is the correctness condition. As for the privacy condition, each user does not get any information about other users' secret messages, either individually in a weak sense or collectively in a perfect sense. 

Such problem is referred as the distributed multi-user secret sharing (DMUSS), which is initialized in \cite{Soleymani2021DistributedMS}. In a DMUSS problem, the decoding collection of each secret message has only one set, which is the subset of the storage nodes that the corresponding user has access to. Also such decoding collection is the same colluding collection of other users' secret messages, either individually for the weak privacy condition or collectively for the perfect privacy condition. The access structure is defined as the union of all decoding collections, which fully characterizes the topology of a DMUSS problem. And we define the secret message rate as the size of a secret message normalized by the size of the share. In \cite{Soleymani2021DistributedMS}, the rate tuple is fixed to be all one and the authors design the access structure and shares. For the perfect privacy condition, \cite{Soleymani2021DistributedMS} proposes to use separate Shamir's scheme \cite{shamir1979how} for each secret message. And \cite{Soleymani2021DistributedMS} has successfully designed a DMUSS scheme with optimal sum of sizes of all shares under the weak privacy condition for some special access structures, the idea is to reuse the same share for multiple secret messages simultaneously.

Later in \cite{Khalesi2021TheCR}, the authors investigated the fundamental limits of the DMUSS problem under the weak privacy condition with arbitrary access structure. They characterize the capacity region given any access structure, defined as the set of all achievable rate tuples, by developing an achievable scheme and proposing a matching converse. The achievable scheme is  still built on individual Shamir's scheme and reuses the shares by a system of linear equations. Compared to \cite{Soleymani2021DistributedMS}, \cite{Khalesi2021TheCR} allows arbitrary access structure and different secret message rates, in particular, it is capacity achieving.

In this paper, we follow the routine of \cite{Khalesi2021TheCR} and investigate the capacity region of the DMUSS problem under the perfect privacy condition. In the converse proof, the essence of the bound is the same as that of \cite{Khalesi2021TheCR}, saying that the size of the secret is upper bounded by the sum of sizes of non-colluding shares. However in the achievability proof, we abandon using Shamir's scheme as a building block and design the encoding function. We have assumed that all shares are mutually independent, and use the fact that the decoding collection of each secret message has only one set, in this sense the decoding function is filled with different indeterminates and zeros and the correctness condition holds. Then it turns out that the perfect privacy condition is equivalent to the full rank property of some matrices extracted from the decoding function. Using the result from the network coding regime, we show that if the field size is bigger than the number of users, a solution does exist. At last, the achievable rate tuple and its converse meet, yielding the capacity region.

The paper is organized as follows. We illustrate the system model in Section II. Section III presents the main result. The converse proof is shown in Section IV and Section V provides the achievability proof. Section VI discusses the weak privacy condition and makes some comparisons. Finally we conclude in Section VII.

\section{system model} \label{sec_sys}
In a three-layer style, the DMUSS problem consists of a master node in the top layer, $ N\in\mathbb{N} $ storage nodes in the middle layer and $ K\in\mathbb{N} $ users in the bottom layer. Secret message $ W_k $ is associated to the $ k $-th user for $ k\in[K]\triangleq\{1,\ldots,K\} $ and all $ K $ secret messages are stored in the master node. The master node has access to all $ N $ storage nodes but it does not have direct access to the users. Storage nodes do not communicate with each other and the share of the $ n $-th storage node is denoted by $ Y_n $ for $ n\in[N] $. For each $ k\in[K] $, the $ k $-th user is connected to a set of storage nodes $ A_k\subseteq[N] $ through some error-free links. The established links are arbitrary, but fixed and known ahead of time to all parties. The set $ A_k $ is referred to as the \emph{access set} for the $ k $-th user and the $ k $-th user can read the entire $ |A_k| $ shares from the corresponding storage nodes. Each storage node is accessed by at least one user and the goal of this problem is to enable each user to retrieve its own secret message correctly and privately from the storage nodes of its access set.

As the topology of the DMUSS problem is mainly about the links between the middle layer and the bottom layer, we define the \emph{access structure} $ \mathcal{A} $ as the collection of all $ K $ access sets:
\begin{align}\label{def_access_structure}
	\mathcal{A}\triangleq\{A_k:k\in[K]\}.
\end{align}
Note that both the number of users $ K $ and the number of storage nodes $ N $ are implied by the access structure $ \mathcal{A} $, i.e., $ K=|\mathcal{A}| $ and $ N=|\cup_{A\in\mathcal{A}}A| $. 

Recall that the access structure can be arbitrary, we assume that each storage node is with unit storage size WLOG. Both the component of the secret message $ W_k $ for the $ k $-th user and the share $ Y_n $ of the $ n $-th storage node are from a finite field $ \mathbb{F}_q $, where  $ q $ is the size of the field. As the size of the secret message may be larger than one, we define the $ k $-th secret message rate as:
\begin{align}\label{def_rate}
	R_k\triangleq H(W_k)/H(Y_n)=H(W_k),
\end{align}
where the base of the logarithm is $ q $. Then we can arrange these $ K $ secret message rates into a vector $ \mathbf{r} $, named as the rate tuple
\begin{align}\label{def_rate_tuple}
	\mathbf{r}\triangleq(R_1,\ldots,R_K).
\end{align}
Moreover, we let $ K $ secret messages be mutually independent and each secret message $ W_k $ is uniformly distributed over $ \mathbb{F}_q^{R_k} $ where $ k\in[K] $ and $ R_k\in\mathbb{N} $, i.e.,
\begin{align}\label{def_independent}
	H(W_1,\hdots,W_K)=\sum_{k\in[K]}H(W_k).
\end{align}

We aim at designing the scheme to work in two phases. In the placement phase, given the $ K $ secret messages, the master node places the encoded $ N $ shares in the corresponding $ N $ storage nodes. In the retrieval phase, for each $ k\in[K] $, the $ k $-th user can successfully reconstruct its designated secret message $ W_k $ by accessing its access set $ A_k $. Let $ Y_{A_k}\triangleq\{Y_i:i\in A_k\} $, we have
\begin{align}\label{def_correctness_condition}
	H(W_k|Y_{A_k})=0,
\end{align}
which is referred as the correctness condition. The conventional notation $ \backslash $ for the difference of two sets will be employed. So the privacy condition, in a \emph{perfect} sense, needs to be satisfied as follows:
\begin{align}\label{def_perfect_condition}
	H(W_{[K]\backslash\{k\}}|Y_{A_k})=H(W_{[K]\backslash\{k\}}),
\end{align}
which means that the $ k $-th user dose not obtain any information, in an information-theoretic sense, about the collection of secret messages from all other users. 
%Also the \emph{weak} privacy condition will be discussed in section.
This scheme can be defined more precisely as follows.
\begin{Definition}[DMUSS Scheme]\label{dmuss_scheme}
	A DMUSS scheme is a bundle of $ (\mathcal{A},\mathbf{r},\mathcal{E},\mathcal{D}) $, where
	\begin{enumerate}
		\item $ \mathcal{A} $ is the access structure as defined in \eqref{def_access_structure}, which characterizes the topology of the DMUSS problem.
		\item $ \mathbf{r} $ is the rate tuple as declared in \eqref{def_rate_tuple}, which indicates the ratio between the size of the secret message and the size of the share. 
%		We assume that $ \mathbf{r}\in\mathbb{N}^K $.
		\item $ \mathcal{E}:\mathbb{F}_q^{\sum_{k\in[K]}R_k}\rightarrow\mathbb{F}_q^N $ is the encoding function that maps the $ K $ secret messages known to the master node to $ N $ shares stored in the $ N $ storage nodes.
		\item $ \mathcal{D} $ is a collection of $ K $ decoding functions $ \mathcal{D}_k: \mathbb{F}_q^{|A_k|}\rightarrow\mathbb{F}_q^{R_k}$, for each $ k\in[K] $, that maps the $ |A_k| $ shares accessed by the $ k $-th user to the designated secret message $ W_k $ correctly. In other words, the correctness condition \eqref{def_correctness_condition} is satisfied.
		\item The perfect privacy condition, as defined in \eqref{def_perfect_condition}, needs to be satisfied.
	\end{enumerate}
\end{Definition}

Following the above preparations, we can define the capacity region as follows:
\begin{Definition}[Capacity Region]
	Given the access structure $ \mathcal{A} $, the capacity region $\mathcal{R}_{\mathcal{A}}$ of this DMUSS problem is the closure of the set of any rate tuple $ \mathbf{r} $ corresponding to a DMUSS scheme $ (\mathcal{A},\mathbf{r},\mathcal{E},\mathcal{D}) $.
\end{Definition}
Note that time sharing can be deployed in the circumstances when the rate tuple contains non-integer ratios. Our achievability proof only provides a linear DMUSS scheme for integral rate tuple.

\section{Main result}\label{main_result}
The main result of the paper is characterizing the capacity region of this DMUSS problem.
\begin{Theorem}
	Given the access structure $ \mathcal{A} $, the capacity region $\mathcal{R}_{\mathcal{A}}$ is a convex region with the non-negative rate tuple $ \mathbf{r} $ satisfying:
		\begin{align}\label{capacity_formula}
			\sum_{i\in S}R_i\leq|\cup_{i\in S}A_i\backslash A_k|,\forall k\in[K],\forall  S\subseteq[K]\backslash\{k\}.
		\end{align}
\end{Theorem}

The converse and achievability proofs are provided in Section \ref{sec_converse} and Section \ref{sec_achievability} respectively. Here we make a few remarks regarding to this theorem.
\begin{Remark}\normalfont
	As can be seen from \eqref{capacity_formula}, if there exists a user $ k $ whose access set has a relatively large number of common elements with the access sets of other users, then the secret message rates of other users are limited to be small. Which can be intuitively explained as the sacrifice for the perfect privacy condition \eqref{def_perfect_condition}. More specifically, though most part of the shares are already known to the $ k $-th user, the $ k $-th user still must not know any information about the secret messages of other users, then the size of those secret messages shall not be big. On the contrary, suppose that the $ k $-th user does not access any share from the storage nodes, i.e., $ A_k=\emptyset $. Then \eqref{capacity_formula} shows that the relation between the size of the secret messages of other users and the size of the corresponding shares can be seen as the  allocation of resource, implied by the mutual independence assumption \eqref{def_independent} and the correctness condition \eqref{def_correctness_condition}.
\end{Remark}
\begin{Remark}\normalfont
	In the converse part, we use a pattern from the single-secret sharing regime to get the same form as \eqref{capacity_formula}. Such pattern is as follows: Given that the shares in the set $ A $ can decode the secret correctly and the shares in the set $ F $ are statistically independent of the secret, then the size of the secret is upper bounded by the sum of the size of the share in the set $ A $ while at the same time not in the set $ F $. Which can be interpreted as that the shares in the intersection of $ A $ and $ F $ contain no information about the secret, then the size of the secret counts on the remaining shares from the set $ A $.
\end{Remark}
\begin{Remark}\normalfont
	In the achievability part, we give an algorithm to show that given the access structure, any integral rate tuple satisfying the capacity formula \eqref{capacity_formula} corresponds to a DMUSS scheme. In this algorithm, we let all shares be mutually independent and the core idea is to design the decoding function of each secret message. The correctness condition follows as the decoding collection for every secret message has only one set. And the perfect privacy condition is transformed into the full rank property of $ K $ matrices consisting of linear combination coefficients and zeros. Such solution exists as long as the rate tuple is bounded by the capacity formula \eqref{capacity_formula} and the field size is larger than $ K $.
\end{Remark}

\section{The converse proof}\label{sec_converse}
We firstly introduce a building block that concludes a relation between the size of the secret and the size of the shares in the single-secret sharing problem.
\begin{Lemma}[Set-Difference Bound]\label{lemma_set_difference} 
	Consider a secret $ W $ and the set of $ N $ shares $ Y_{[N]} $, let the decoding collection $ \mathcal{A} $ contains every subset of $ Y_{[N]} $ such that each can successfully decode the secret $ W $, similarly the colluding collection $ \mathcal{F} $ contains every subset of $ Y_{[N]} $ such that each knows no information about the secret $ W $. Then for any element $ A\in\mathcal{A} $ and $ F\in\mathcal{F} $,
	\begin{align}
		H(W)\leq\sum_{i\in A\backslash F}H(Y_i).
	\end{align}
\end{Lemma}
\begin{IEEEproof}
	\begin{align}
		\label{decodable_secure}H(W)&=H(W|Y_F)-H(W|Y_A)\\
		\nonumber&=H(W|Y_F)-H(W|Y_{A\cup F})\\
		\nonumber&=I(W;Y_{A\backslash F}|Y_F)\\
		\nonumber&\leq H(Y_{A\backslash F}|Y_F)\\
		\nonumber&\leq H(Y_{A\backslash F})\\
		&\leq\sum_{i\in A-F}H(Y_i)\nonumber
	\end{align}
\end{IEEEproof}
Note that \eqref{decodable_secure} is based on the correctness condition that $ H(W|Y_A)=0 $ and the privacy condition that $ H(W|Y_F)=H(W) $, the others are by nothing but the properties of conditional entropy and conditional mutual information. 
\begin{Remark}\normalfont
	In fact, this bound shares the same idea with Cheng \cite[Lemma 2]{cheng2014performance} as he considered arbitrary wiretap sets given a general wiretap network.
	We also believe this bound is already known to the secret sharing community as it conveys a physical meaning that the size of the secret is upper bounded by the sum of sizes of non-colluding shares. 
	In the single-secret sharing regime, there indeed exist other more complicated patterns as discovered in \cite{stinson1992an,jackson1996perfect,9127978}, since the decoding collection $ \mathcal{A} $ for correctness and the colluding collection $ \mathcal{F} $ for privacy can very a lot.
\end{Remark}

Based on Lemma \ref{lemma_set_difference}, for each $ k\in[K] $ and any non-empty subset $ S\subseteq[K]\backslash\{k\}$, we have
\begin{align}\label{converse_1}
	H(W_S)\leq\sum_{j\in\cup_{i\in S}A_i\backslash A_k}H(Y_j),
\end{align}
as the shares of the set $ \cup_{i\in S}A_i $ can decode the collection of secret messages $ W_S $ by the correctness condition \eqref{def_correctness_condition} and the shares in the access set $ A_k $ does not know any information about $ W_S $ by the perfect privacy condition \eqref{def_perfect_condition}. Finally, recall that the $ K $ secret messages are mutually independent, each share is with unit size and by the definition of secret message rate in \eqref{def_rate}, we can rewrite \eqref{converse_1} as
\begin{align}
	\sum_{i\in S}R_i\leq|\cup_{i\in S}A_i\backslash A_k|,
\end{align}
which matches with the capacity formula \eqref{capacity_formula} to characterize the capacity region.

\section{The achievability proof}\label{sec_achievability}
In this section, we firstly introduce the linear scheme for single-secret sharing problem, where an example whose decoding collection has only one set is illustrated. 
Similar to this, we give the linear DMUSS scheme and discuss its feasibility. Then it turns out that given the access structure, any integral rate tuple satisfying the capacity formula \eqref{capacity_formula} corresponds to a feasible DMUSS scheme, thus complete the achievability proof.

\subsection{Single-secret sharing scheme}
Before introducing the so-called \textit{linear scheme} \cite{karnin1983secret}, some concepts need to be illustrated.  Let $\mathcal{V}$ be a vector space with finite dimension $n$ over a finite field $\mathbb{F}_q$. And $ \mathcal{O} $ is the set of random variables including the secret and shares. For every $i\in\mathcal{O}$, denote a subset of $\mathcal{V}$ by ${V}_i$, we assume that all vectors of ${V}_i$ are linear independent, then the length of such subset is $ \text{rk}(V_i) $, here $ \text{rk}(.) $ calculates the rank of its input. Hereafter, we treat each ${V}_i$ as a matrix of size $ n\times\text{rk}(V_i)$ and stack such $ |\mathcal{O}| $ matrices in sequence horizontally (column wise) into a bigger matrix $\mathbf{G}$ of size $ n\times\sum_{i\in\mathcal{O}}\text{rk}({V}_i) $, referred as a generator matrix. 

Then consider a uniform discrete probability distribution whose alphabet is the set of $q^n$ different row vectors with length $n$ taking values from $\mathbb{F}_q$. These vectors can be stacked vertically into a matrix $\mathbf{C}\in \mathbb{F}_q^{q^n\times n}$. For $\mathbf{CG}$, with the set of row vectors as the alphabet and that the probability of each row equals $1/q^n$, it forms a new discrete probability distribution with random variables from the set $\mathcal{O}$, where 
\begin{align}\label{rank_entropy}
	H(X)=\text{rk}({V}_X),\forall\emptyset\neq X\subseteq\mathcal{O},
\end{align} 
with the base of the logarithm being $ q $ and $ {V}_X $ denotes a matrix stacked horizontally by the corresponding $ |X| $ matrices. Note that if $\text{rk}({V}_{\mathcal{O}})<n$, then some row vectors of $\mathbf{CG}$ are identical and the corresponding probabilities need to be summed, otherwise, all rows are unique.

A generator matrix satisfying certain correctness condition and privacy condition can be referred as a linear secret sharing scheme. For example consider a secret $ W $ and the set of $ 3 $ shares $ Y_{[3]} $, let the decoding collection $ \mathcal{A}$ contains $ Y_{[N]} $ only, and the colluding collection $ \mathcal{F} $ contains every subset of $ Y_{[N]} $ whose cardinality equals two. In other words, we target on a threshold single-secret sharing problem with only one decoding choose. A linear scheme can be determined by the following generator matrix over $ \mathbb{F}_5 $:
\begin{equation}\label{single_enc}
	\left[ {\begin{array}{c|c|c|c}
			\begin{matrix}
				1 \\
				0 \\
				0
			\end{matrix}
			&
			\begin{matrix}
				1\\
				1\\
				1
			\end{matrix}
			&
			\begin{matrix}
				1\\
				2\\
				2^2
			\end{matrix}
			&
			\begin{matrix}
				1\\
				3\\
				3^2
			\end{matrix}
	\end{array}} \right].
\end{equation}
Note that every codeword corresponds to a distribution of shares as mentioned in \cite[Section VI]{9127978}. The vertical bars indicate which positions of the codeword correspond to the secret and to every share. In this case, a codeword
\begin{align}
	\begin{bmatrix}
		w & y_1 & y_2 & y_3
	\end{bmatrix}\in\mathbb{F}_5^4
\end{align}
corresponds to a distribution of shares where the secret value is $ w $, the first share is $ y_1 $, and so on. The correctness condition follows by the full rank property of the Vandermonde matrix in \eqref{single_enc}, i.e., $ \text{rk}(V_{W,Y_{[3]}})=\text{rk}(V_{Y_{[3]}}) $. As for the privacy condition, any sub-matrix of size $ 2\times2 $ from $ V_{Y_{[3]}} $ in the bottom is full rank, thus we have $ \text{rk}(V_{W,Y_{\{i,j\}}})=\text{rk}(V_{Y_{\{i,j\}}})+\text{rk}(V_{W}) $, where $ i,j\in[3],i\neq j $.

Similar to DMUSS scheme as mentioned in Definition \ref{dmuss_scheme}, single-secret sharing scheme also works in placement and retrieval phases. Still using the above example, in the placement phase, we firstly introduce two random keys $ K_1 $ and $ K_2 $ drawn independently and uniformly from $ \mathbb{F}_5 $ like the secret $ W $. Then the dealer arrange the secret and these two keys into a row vector. Finally, for each $ i\in[3] $, the share $ Y_i $ equals this row vector multiplied by the column vector $ V_{Y_i} $ from the generator matrix \eqref{single_enc}. As we can see, the column vector $ V_{Y_i}  $ corresponds essentially to the {encoding function} for the share $ Y_i $. In fact, this liner scheme is equivalent to Shamir's scheme \cite{shamir1979how}, which is based on polynomial evaluation.

After the above declaration of placement phase and from the generator matrix \eqref{single_enc}, we can interpret the privacy condition of this single-secret sharing scheme as that the two random keys protect the secret by a careful design of the linear combination coefficients for the share. More specifically, any two shares know nothing about the secret since the information of two random keys can not be removed from the secret by the full rank property of the sub-matrix at the bottom of $ V_{Y_{[3]}} $. The correctness condition follows because three shares can figure out any random key and the secret by the full rank property of the Vandermonde matrix.

In this example, these three shares are mutually independent still by the full rank property of the Vandermonde matrix, and the number of random keys plus the size of the secret equals the number of shares exactly. Utilizing Gaussian-Elimination towards the matrix $ V_{Y_{[3]}} $, we can rewrite the generator matrix \eqref{single_enc} as follows:
\begin{equation}\label{single_dec}
	\left[ {\begin{array}{c|c|c|c}
			\begin{matrix}
				3 \\
				2 \\
				1
			\end{matrix}
			&
			\begin{matrix}
				1\\
				0\\
				0
			\end{matrix}
			&
			\begin{matrix}
				0\\
				1\\
				0
			\end{matrix}
			&
			\begin{matrix}
				0\\
				0\\
				1
			\end{matrix}
	\end{array}} \right],
\end{equation}
which generates the same $ 5^3 $ codewords as \eqref{single_enc}. The benefit of this form is that the first column vector corresponds essentially to the {decoding function} for the secret. More specifically, if we arrange the three shares into a row vector, finally the secret equals such row vector multiplied by the first column vector of \eqref{single_dec}, which completes the retrieval phase.

From the generator matrix \eqref{single_dec}, the column vectors corresponding to all shares actually form an identity matrix, which means that all shares together know the full information of both the secret and the random keys. In this way, we can treat each share as a random key and the secret is fixed to be a linear combination of these keys, which is reflected in the column vector(s) corresponding to the secret, so the correctness condition immediately follows. As for the privacy condition, in the above example when we verify that two shares $ Y_1 $ and $ Y_2 $ are statistically independent of the secret, which can also be seen as that the secret knows nothing about these two shares. Then the share $ Y_3 $ plays the role to protect the other two shares. As can be seen from the first column vector, the third component is non-zero, that is, the secret can not remove the information of $ Y_3 $ from $ Y_1 $ and $ Y_2 $, finally the privacy condition holds.

\subsection{DMUSS scheme}
Our proposed DMUSS scheme is summarized in the following algorithm, and we will discuss its feasibility in detail.

\begin{algorithm}
%	\caption{decoding design}
	\caption{}
	\algsetup{linenodelimiter=.}
	\begin{algorithmic}[1]\label{algorithm}
		 \REQUIRE An access structure $ \mathcal{A} $ and an integral rate tuple $ \mathbf{r} $ satisfying the capacity formula \eqref{capacity_formula}.
		 \ENSURE A generator matrix of size $ N\times(\sum_{k\in[K]}R_k+N) $.
		\STATE Fix $ V_{Y_{[N]}} $ to be an identity matrix of size $ N\times N $. 
		\STATE Let $ V_{W_{[K]}} $ be full of zeros and different indeterminates. More specifically, for each $ k\in[K] $, let $ V_{W_k} $ be a size $ N\times R_k $ matrix, where each component in the rows with indices in the access set $ A_k $ is an indeterminate and the other rows are all zero.
		\STATE  For each $ k\in[K] $,  find $ K-1 $ subsets $ C_i^k\subseteq[N] $, $ i\in[K]\backslash\{k\} $, satisfying the following three conditions:
		\begin{align}
			C_i^k&\subseteq A_i\backslash A_k,\label{subset_}\\
			|C_i^k|&=R_i,\label{size_}\\
			C_i^k&\cap C_{i'}^k=\emptyset,\forall i\neq i'.\label{mutual_}
		\end{align}
%		\begin{enumerate}
%			\item \begin{align}
%				23
%			\end{align}
%			\item 4324
%		\end{enumerate}
		\STATE For each $ k\in[K] $, from the matrix $ V_{W_{[K]\backslash\{k\} }} $ extract a size $ \sum_{i\in[K]\backslash\{k\} }R_i\times \sum_{i\in[K]\backslash\{k\} }R_i $ sub-matrix with row indices in the set $ \cup_{i\in[K]\backslash\{k\}}C_i^k $, finally denote such square sub-matrix as $ E_k $.
		\STATE Let $ \text{det}(.) $ calculate the determinant of its input square matrix. Once we find a solution to the nonzero polynomial
		\begin{align}
			\prod_{k\in[K]}\text{det}(E_k)\neq0,
		\end{align}
		and let other uninvolved indeterminates be zero, we refer this generator matrix as a DMUSS scheme.
	\end{algorithmic}
\end{algorithm}

\subsubsection{Input and Output}
Recall that the access structure of the DMUSS problem is the collection of access sets from all users and the rate tuple encapsulates the size of the secret message normalized by the size of the share. Given an access structure and a rate tuple as input to the algorithm, we need to design the encoding and decoding functions to build a DMUSS scheme, such functions will be implied by the generator matrix as the output of this algorithm.

\subsubsection{Initialization}
In this algorithm, we have assumed that all shares are mutually independent, so the matrix $ V_{Y_{[N]}} $ corresponding to all $ N $ shares is fixed to be an identity matrix.
For each $ k\in[K] $, we let the matrix $ V_{W_k} $ correspond to the decoding function for the secret message $ W_k $, that is, $ W_k $ equals the row vector formed by all shares multiplied by $ V_{W_k} $. In a general sense, we put different indeterminates in the rows with indices in the access set $ A_k $, the other rows are fixed to be zero since the decoding collection of each secret message has only one set as indicated in the correctness condition \eqref{def_correctness_condition}. 
%the shares from the remaining set $ [N]\backslash A_k $ are not needed in the decoding phase.

\subsubsection{The Perfect Privacy Condition}
Note that after the initialization in steps 1 and 2, only the perfect privacy condition \eqref{def_perfect_condition} remains to be investigated. 
Recall that for each $ k\in[K] $, such condition asks that the shares from the storage nodes corresponding to the access set $ A_k $ are statistically independent of the secret messages from all other users.
Similar to the generator matrix \eqref{single_dec}, we treat each share as a random key and interpret the perfect privacy condition as that the collection of the secret messages $ W_{[K]\backslash\{k\}} $ knows no information about the shares $ Y_{A_k} $. 
In this way, the other shares $ Y_{[N]\backslash A_k} $ play the role to protect $ Y_{A_k} $.
In order to achieve this goal, the corresponding linear combination coefficients need to be designed carefully so that the secret messages $ W_{[K]\backslash\{k\}} $ can not remove the information of $ Y_{[N]\backslash A_k} $ from $ Y_{A_k} $. 
More specifically, for each $ k\in[K] $, let $ E_k' $ denote the matrix extracted from the matrix $ V_{W_{[K]\backslash\{k\} }} $ with row indices in the set $ [N]\backslash A_k $, finally the perfect privacy condition \eqref{def_perfect_condition} is equivalent to
\begin{align}\label{ek_prime}
	\text{rk}(V_{W_{[K]\backslash\{k\} }})=\text{rk}(E_k'),
\end{align}
since the equivalence between rank and entropy as in \eqref{rank_entropy} and that $ V_{Y_{[N]}} $ is an identity matrix. 

Coupled with the fact that \eqref{ek_prime} holds if and only if there exists a $ \sum_{i\in[K]\backslash\{k\} }R_i\times\sum_{i\in[K]\backslash\{k\} }R_i $ sub-matrix of the matrix $ E_k' $ with nonzero determinant. And as such sub-matrix contains the mixture of indeterminates and zeros, the nonzero determinant can be seen as a polynomial function taking nonzero value. Hence, two circumstances can be developed:
\begin{enumerate}
	\item At least one coefficient in the polynomial is nonzero.
	\item Choose scalar values for the indeterminates so that the polynomial function assumes a nonzero scalar value.
\end{enumerate} 
Note that if the first circumstance fails, no matter what scalar values for the indeterminates we choose, the polynomial function is fixed to be zero.
By the calculation of determinant, the first circumstance asks a special position of the indeterminates in a square matrix, this goal is pursued by steps 3 and 4 of the achievability algorithm. And step 5 corresponds to the second circumstance. We discuss these in detail as follows.

\subsubsection{The first circumstance}
The determinant of a $ l\times l $ matrix can be defined by Leibniz formula, which is a sum of signed products of matrix entries such that each summand is the product of $ l $ different entries, which means that every column of the matrix corresponds to an entry and all $ l $ entries are in different rows, so the number of these summands is $ l! $. Recall that an entry of any sub-matrix of the matrix $ E_k' $ is either a zero or a unique indeterminate. Then if there exists a sub-matrix such that one of the summands is the product of $ l $ indeterminates, we have that at least one coefficient in the polynomial is nonzero. 

Such assumption is satisfied according to the three conditions for finding $ K-1 $ row index sets $ C_i^k $ in step 3 of the algorithm. More specifically, the first condition \eqref{subset_} delimits the boundary, that is, in the matrix $ V_{W_i} $ indeterminates with row indices $ A_i\backslash A_k $ are ready to form a summand. And from the matrix $ V_{W_i} $ choose $ R_i $ number of indeterminates, which is also the number of columns of $ V_{W_i} $, according to the second condition \eqref{size_}. Finally the third condition \eqref{mutual_} guarantees non-overlapping of the indeterminates, thus the determinant of such sub-matrix $ E_k $ has at least one summand with $ \sum_{i\in[K]\backslash\{k\} }R_i $ different indeterminates. 
%An example regarding to such summand part will be presented in the next subsection.

\subsubsection{Existence of $ C_i^k $}
Before introducing the existence of these $ K-1 $ row index sets $ C_i^k $, we firstly borrow a theorem \cite{Cameron1995CombinatoricsTT} from combinatorial mathematics.
\begin{Theorem}[Hall's Marriage Theorem]
	Let $ \mathcal{B} $ be a collection of finite sets and $ \mathcal{B} $ can contain the same set multiple times. A sufficient and necessary condition for being able to select a distinct element from each finite set is that $ \mathcal{B} $ satisfies the marriage condition:
	\begin{align}\label{marriage_condition}
		|\mathcal{G}|\leq |\cup_{U\in\mathcal{G}}U|,\forall\mathcal{G}\subseteq\mathcal{B}.
	\end{align}
\end{Theorem}
Note that the marriage condition asks that each subfamily $ \mathcal{G} $ contains at least as many distinct members as its number of sets. 

In the following, we will discuss the relation between the three conditions \eqref{subset_}-\eqref{mutual_} and the extracting distinct elements problem. To begin with, we duplicate the set $ A_i\backslash A_k $ $ R_i $ times. More specifically, let
\begin{align}\label{duplicate_set}
	B_{i,j}^k\triangleq A_i\backslash A_k,j\in[R_i].
\end{align}
Then form a collection $ \mathcal{B}^k $ of all $ B_{i,j}^k, i\in[K]\backslash\{k\},j\in[R_i] $. It turns out that if we can extract a distinct element $ b_{i,j}^k $ from each set $ B_{i,j}^k $ of $ \mathcal{B}^k $, with
\begin{align}\label{build_c}
	C_i^k=\{b_{i,j}^k:j\in[R_i]\},
\end{align}
the first condition \eqref{subset_} follows by the initialization of the set $ B_{i,j}^k $, the second condition \eqref{size_} holds since the $ R_i $ elements are distinct and it is the same reason for the third condition \eqref{mutual_}. 

To justify that we can successfully extract distinct elements, we need to test whether such $ \sum_{i\in[K]\backslash\{k\} }R_i $ sets $ B_{i,j}^k $ satisfy the marriage condition \eqref{marriage_condition} according to Hall's Marriage Theorem. Then for any subfamily $ \mathcal{G}\subseteq\mathcal{B}^k $, let the set $ S $ contain all indices $ i $ such that there exists an index $ j\in[R_i] $ with $ B_{i,j}^k\in\mathcal{G} $. So we have
\begin{align}\label{marriage_1}
	|\mathcal{G}|\leq\sum_{i\in S}R_i,
\end{align}
since for each $ i\in S $ there are at most $ R_i $ same sets $ A_i\backslash A_k $.
Recall the formula \eqref{capacity_formula} of the capacity region, for fixed $ k\in[K] $ we have upper bounded the secret message rates:
\begin{align}\label{marriage_2}
	\sum_{i\in S}R_i\leq|\cup_{i\in S}A_i\backslash A_k|.
\end{align}
The RHS is equivalent to $ |\cup_{U\in\mathcal{G}}U| $ due to the copy nature in building each set $ B_{i,j}^k $. Finally combine \eqref{marriage_1} and \eqref{marriage_2} together, the marriage condition follows. So distinct elements can be extracted, furthermore, we claim the existence of $ K-1 $ row index sets $ C_i^k $ by \eqref{build_c}.

To conclude the above existence proof, we firstly build an extracting distinct elements problem \eqref{duplicate_set} to imply the three conditions \eqref{subset_}-\eqref{mutual_} of $ C_i^k $, then by Hall's Marriage Theorem, solving such problem depends on the marriage condition \eqref{marriage_condition}, which is satisfied exactly by the capacity formula \eqref{capacity_formula}.

As for practical implementation of step 3, for each $ k\in[K] $, the extracting distinct elements problem is equivalent to solving a bipartite matching or a max-flow optimization problem with $ N+\sum_{i\in[K]\backslash\{k\} }R_i\leq 2N $ vertices and $ \sum_{i\in[K]\backslash\{k\} }R_i|A_i\backslash A_k|\leq(\sum_{i\in[K]\backslash\{k\} }R_i)N\leq N^2  $ edges. Hopcroft and Karp algorithm \cite[p. 763]{Cormen1990IntroductionTA} can be used to obtain such a matching graph with computation complexity of $ O(\sqrt{2N}N^2)=O(N^{2.5}) $.

\subsubsection{The second circumstance}
So from steps 3 and 4 of this algorithm, the perfect privacy condition \eqref{def_perfect_condition} is equivalent to the full rank property of $ K $ square matrices $ E_k $, and the determinant of each matrix is a polynomial with at least one nonzero coefficient. We multiply these $ K $ polynomials together, and the existence of a solution for indeterminates to make the final polynomial nonzero is supported by the following lemma \cite[Lemma 19.17]{Yeung2008InformationTA}.

\begin{Lemma}
	Let $ g(z_1,\ldots,z_n) $ be a nonzero polynomial with coefficients in a field $ \mathbb{F}_q $. If $ q $ is greater than the degree of $ g $ in every $ z_i $, then there exist $ a_1,\ldots,a_n\in\mathbb{F}_q $ such that $ g(a_1,\ldots,a_n)\neq0 $.
\end{Lemma}
Note that in the regime of network coding, an efficient single-source multi-cast solution also needs the full rank property of certain collections of global encoding kernels. This lemma shows that when provided with an enough large field, such solution does exist. As each summand of the determinant of the matrix $ E_k $ is the product of $ \sum_{i\in[K]\backslash\{k\} }R_i $ different indeterminates, the degree of a single determinant is one. Then step 5 asks these $ K $ matrices to be full rank simultaneously, the degree of the final polynomial is at most $ K $. Thus we have $ q>K $.

For practical implementation of step 5, we still borrow an algorithm \cite[Algorithm 1]{Koetter2001AnAA} from network coding regime to give an explicit solution. The core idea behind this algorithm is that it assigns the intermediates to scalar values from the field one by one, as long as the intermediate polynomial, after assignment, is nonzero, i.e., the intermediate polynomial has at least one nonzero coefficient. Such assignment of an intermediate is guaranteed as the size of the field is larger than the degree of the polynomial. When this field size assumption is broken, however, some intermediate polynomial may be zero no matter what scalar value to choose. The computation complexity is $ O(q^n) $, where $ q $ is the field size and $ n $ is the number of total intermediates.

When the field size is allowed to be large, we can randomly choose the scalar values and with high probability, the polynomial is nonzero. More specifically, let $ m $ be the highest degree of polynomial $ g $ in $ z_i,i\in[n] $. Then choose $ a_1,\ldots,a_n $ independently according to the uniform distribution on the field $ \mathbb{F}_q $, the probability of a nonzero polynomial \cite[Corollary 19.18]{Yeung2008InformationTA} is at least
\begin{align}
	(1-m/q)^n.
\end{align}
Note that the larger field size, the higher probability.

\subsubsection{Decoding and encoding functions}
After step 5, we have determined all intermediates and the final generator matrix shows all $ K $ decoding functions $ V_{W_{[K]}} $ regarding to the linear combinations of shares. As for the encoding function, similar to the relation between the generator matrix \eqref{single_enc} for encoding function and the generator matrix \eqref{single_dec} for decoding function, we use Gaussian-Elimination to the entire size $ N\times(\sum_{k\in[K]}R_k+N) $ generator matrix to get a new one in row echelon form. In particular, the new matrix $ V_{W_{[K]}} $ is an identity matrix stacked with an all zero matrix vertically. In this way, for each $ n\in[N] $, the new matrix $ V_{Y_n} $ corresponds to the encoding function for the share $ Y_n $. More specifically, $ Y_n $ equals the row vector formed by all secret messages and $ N- \sum_{k\in[K]}R_k$ random keys multiplied by the matrix $ V_{Y_n} $. The computation complexity behind this transform is from the Gaussian-Elimination, which is $ O(N^3) $.

\subsubsection{Mutual Independence Assumption of Secret Messages}
Recall that we have assumed all $ K $ secret messages are mutually independent in \eqref{def_independent}, one fact is that as long as the DMUSS scheme satisfies the correctness condition \eqref{def_correctness_condition} and the perfect privacy condition \eqref{def_perfect_condition}, all secret messages have to be mutually independent. In other words, if the secret messages are correlated, the DMUSS scheme shall not exist.

The reason is that, for example, if the correctness condition $ H(W_1|Y_{A_1})=0 $ and the prefect privacy condition $ H(W_{[K]\backslash\{1\}}|Y_{A_1})=H(W_{[K]\backslash\{1\}}) $ hold, then we have
\begin{align}
	H(W_{[K]\backslash\{1\}}|W_1)=H(W_{[K]\backslash\{1\}}).
\end{align}
There are total $ K $ such formulas, and finally we have the same equation \eqref{def_independent}.

So in the algorithm, when the correctness condition is already satisfied and the perfect privacy condition is pursued, the final DMUSS scheme has the property that $ K $ secret messages are mutually independent. More specifically, one can also check with the formula \eqref{ek_prime}, which builds a bridge between the perfect privacy condition and the full rank property of a certain matrix.

\subsection{An example}\label{sec_example}
Consider an example with four users and six storage nodes, the topology is reflected in the access structure $ \mathcal{A} $ consisting of four access sets:
\begin{align}
	A_1=\{1,2,4\},A_2=\{2,3,6\},A_3=\{1,4,5\},A_4=\{3,5,6\}.
\end{align}
Then based on the capacity formula \eqref{capacity_formula}, we can let the rate tuple be
\begin{align}
	\mathbf{r}=(1,1,1,1).
\end{align}

After the above input to the algorithm, we initialize the size $ 6\times10 $ generator matrix as follows:
\begin{align}
	\begin{bmatrix}
		d_{1,1} & 0 & d_{3,1} & 0 & 1 & 0 & 0 & 0 & 0 & 0 \\
		d_{1,2} & d_{2,2} & 0 & 0 & 0 & 1 & 0 & 0 & 0 & 0 \\
		0 & d_{2,3} & 0 & d_{4,3} & 0 & 0 & 1 & 0 & 0 & 0 \\
		d_{1,4} & 0 & d_{3,4} & 0 & 0 & 0 & 0 & 1 & 0 & 0 \\
		0 & 0 & d_{3,5} & d_{4,5} & 0 & 0 & 0 & 0 & 1 & 0 \\
		0 & d_{2,6} & 0 & d_{4,6} & 0 & 0 & 0 & 0 & 0 & 1
	\end{bmatrix}.
\end{align}
Note that similar to the generator matrix \eqref{single_dec}, the column vector $ V_{W_k} $ corresponds to the decoding function of the secret message $ W_k $ and the indeterminates are aligned with the access set $ A_k $.

Then in step 3, we can find the row index sets $ C_i^k $ as follows:
\begin{align}
	C_2^1=\{3\},C_3^1=\{5\},C_4^1=\{6\},\nonumber\\
	C_1^2=\{1\},C_3^2=\{4\},C_4^2=\{5\},\nonumber\\
	C_1^3=\{2\},C_2^3=\{3\},C_4^3=\{6\},\nonumber\\
	C_1^4=\{1\},C_2^4=\{2\},C_3^4=\{4\}.\nonumber
\end{align}
It can be checked that these sets satisfy the three conditions \eqref{subset_}-\eqref{mutual_}. Then after step 4, we have $ 4 $ sub-matrices $ E_k $ such that the perfect privacy condition is equivalent to the full rank property of these matrices. We list these matrices as follows:
\begin{align}
	E_1=\begin{bmatrix}
		d_{2,3} & 0 & d_{4,3} \\
		0 & d_{3,5} & d_{4,5} \\
		d_{2,6} & 0 & d_{4,6}
	\end{bmatrix},	E_2=\begin{bmatrix}
	d_{1,1} & d_{3,1} & 0 \\
	d_{1,4} & d_{3,4} & 0 \\
	0 & d_{3,5} & d_{4,5}
\end{bmatrix},	\nonumber\\
E_3=\begin{bmatrix}
d_{1,2} & d_{2,2} & 0 \\
0 & d_{2,3} & d_{4,3} \\
0 & d_{2,6} & d_{4,6}
\end{bmatrix},	E_4=\begin{bmatrix}
d_{1,1} & 0 & d_{3,1} \\
d_{1,2} & d_{2,2} & 0 \\
d_{1,4} & 0 & d_{3,4}
\end{bmatrix}.\nonumber
\end{align}

It can be seen that the determinant of each matrix is a nonzero polynomial, i.e., at least one coefficient is nonzero. Hence in step 5, in order to make these four matrices full rank simultaneously, we need to find a solution to the following nonzero polynomial, which is the multiplication of these four determinants:
\begin{align}
	(d_{2,3}d_{3,5}d_{4,6}-d_{2,6}d_{3,5}d_{4,3})(d_{1,1}d_{3,4}d_{4,5}-d_{1,4}d_{3,1}d_{4,5})\nonumber\\
	(d_{1,2}d_{2,3}d_{4,6}-d_{1,2}d_{2,6}d_{4,3})(d_{1,1}d_{2,2}d_{3,4}-d_{1,4}d_{2,2}d_{3,1})\neq0.
\end{align}
Then we give a solution from $ \mathbb{F}_2 $ as follows:
\begin{align}
	\begin{bmatrix}
		1 & 0 & 0 & 0 & 1 & 0 & 0 & 0 & 0 & 0 \\
		1 & 1 & 0 & 0 & 0 & 1 & 0 & 0 & 0 & 0 \\
		0 & 1 & 0 & 0 & 0 & 0 & 1 & 0 & 0 & 0 \\
		0 & 0 & 1 & 0 & 0 & 0 & 0 & 1 & 0 & 0 \\
		0 & 0 & 1 & 1 & 0 & 0 & 0 & 0 & 1 & 0 \\
		0 & 0 & 0 & 1 & 0 & 0 & 0 & 0 & 0 & 1
	\end{bmatrix}.
\end{align}
According to this result, the decoding function of the secret message $ W_1 $ is $ Y_1+Y_2 $, and so on.

At last we apply Gaussian-Elimination towards above generator matrix and we have:
\begin{align}
	\begin{bmatrix}
		1 & 0 & 0 & 0 & 1 & 0 & 0 & 0 & 0 & 0 \\
		0 & 1 & 0 & 0 & 0 & 0 & 1 & 0 & 0 & 0 \\
		0 & 0 & 1 & 0 & 0 & 0 & 0 & 1 & 0 & 0 \\
		0 & 0 & 0 & 1 & 0 & 0 & 0 & 0 & 0 & 1 \\
		0 & 0 & 0 & 0 & 1 & 1 & 1 & 0 & 0 & 0 \\
		0 & 0 & 0 & 0 & 0 & 0 & 0 & 1 & 1 & 1
	\end{bmatrix}.
\end{align}
The encoding function can be seen with the help of two random keys $ K_1 $ and $ K_2 $, according to this result, the encoding function of the share $ Y_1 $ is $ W_1+K_1 $, the share $ Y_4=W_3+K_2 $, and so on.

% \section{Extension to Arbitrary Collusion Pattern}
\section{The Weak Privacy Condition}
In this section, we give the weak privacy condition, discuss the corresponding capacity region and compare the existing achievability proof \cite[Section IV]{Khalesi2021TheCR} and our proposed algorithm.

\subsection{The Capacity Region}
Recall that the perfect privacy condition \eqref{def_perfect_condition} asks that the shares $ Y_{A_k} $ of the $ k $-th user are statistically independent of secret messages $ W_{[K]\backslash\{k\}} $ from all other users, while the weak privacy condition only focuses on the individual secret message $ W_{k'} $ from any other user. More specifically, for any $ k,k'\in[K],k\neq k' $, we have
\begin{align}\label{weak_privacy}
	H(W_{k'}|Y_{A_k})=H(W_{k'}).
\end{align}
Such weak privacy condition is also employed in regenerating code \cite{Bhattad2005WeaklySN} and secure network coding \cite{Kadhe2014WeaklySR}. 

For the DMUSS problem, \cite{Khalesi2021TheCR} has already characterized the capacity region under the weak privacy condition \eqref{weak_privacy} as follows:
\begin{Theorem}
	Given the access structure $ \mathcal{A} $, when under the weak privacy condition, the capacity region $\mathcal{R}_{\mathcal{A}}^w$ is a convex region with the non-negative rate tuple $ \mathbf{r} $ satisfying:
	\begin{align}
      R_k & \leq \min_{k  \neq {k'}}|A_k\backslash A_{{k'}}|,\forall{k,{k'}\in[K]},\label{weak_capacity_1}\\
\sum_{i\in S}R_i & \leq|\cup_{i\in S} A_i|,\forall S\subseteq[K].
\label{weak_capacity_2}
	\end{align}
\end{Theorem}
\begin{Remark}\normalfont
	In the converse proof, similar to the case when the perfect privacy condition is considered, we can use the same pattern that the size of the secret is upper bounded by the sum of sizes of non-colluding shares, to get \eqref{weak_capacity_1} from the correctness condition \eqref{def_correctness_condition} and the weak privacy condition \eqref{weak_privacy}. 
	As the perfect privacy condition is stronger than and imply the weak privacy condition, we can see that the bound \eqref{weak_capacity_1} is contained in \eqref{capacity_formula}. As for \eqref{weak_capacity_2}, it  actually means the resource allocation, implied by the mutual independence assumption of secret messages \eqref{def_independent} and the correctness condition \eqref{def_correctness_condition}.
\end{Remark}
\begin{Remark}\normalfont
	In the achievability proof, \cite{Khalesi2021TheCR} designs the encoding function based on individual Shamir's scheme for each secret message, such that the share which is accessed by multiple users is actually reused, which is guaranteed by a system of linear equations. We will discuss this method without proof via an example, mainly focus on the idea.
%	and the comparison with our proposed algorithm for designing the decoding function.
\end{Remark}

\subsection{An Example}
Consider an example with two users and four storage nodes, the access structure consists of two access sets:
\begin{align}
	A_1=\{1,2,3\},A_2=\{3,4\},
\end{align}
and the rate tuple is an integral vector satisfying \eqref{weak_capacity_1} and \eqref{weak_capacity_2}:
\begin{align}
	\mathbf{r}=(2,1).
\end{align}

When consider two individual Shamir's schemes independently, we can build the encoding functions with the help of two random keys $ K_1$ and $ K_2 $ as follows:
\begin{align}
	\begin{bmatrix}
		1 & 1 & 1 \\
		1 & 2 & 2^2 \\
		1 & 3 & 3^2
	\end{bmatrix}&\cdot\begin{bmatrix}
	W_{1,1} \\
	W_{1,2} \\
	K_1
\end{bmatrix}=\begin{bmatrix}
Y_1 \\
Y_2 \\
Y_3
\end{bmatrix},\\
	\begin{bmatrix}
	1 & 1  \\
	1 & 2 
\end{bmatrix}&\cdot\begin{bmatrix}
	W_{2} \\
	K_2
\end{bmatrix}=\begin{bmatrix}
	Y_3 \\
	Y_4
\end{bmatrix}.
\end{align}
In this way, the size of the share $ Y_3 $ has to be two since the 3-rd storage node is accessed by two users at the same time. In order to be efficient, \cite{Khalesi2021TheCR} uses the idea to reuse the share $ Y_3 $ for both secret messages simultaneously, by jointly combining these two Shamir's schemes through a system of linear equations. The detail is as follows:
\begin{align}\label{initial_system_linear}
	\begin{bmatrix}
		1 & 1 & 1    & 0 & 0 & \alpha_{1,1} & 0 & 0 & 0 \\
		1 & 2 & 2^2 & 0 & 0 & 0 & \alpha_{1,2} & 0 & 0 \\
		1 & 3 & 3^2 & 0 & 0 & 0 & 0 & \alpha_{1,3} & 0 \\
		0 & 0 & 0 & 1 & 1 & 0 & 0 & \alpha_{2,3} & 0 \\
		0 & 0 & 0 & 1 & 2 & 0 & 0 & 0 & \alpha_{2,4}
	\end{bmatrix}\cdot\begin{bmatrix}
	W_{1,1} \\
	W_{1,2}\\
	K_1\\
	W_{2}\\
	K_2\\
	Y_1\\
	Y_2\\
	Y_3\\
	Y_4
\end{bmatrix}=\begin{bmatrix}
0 \\
0 \\
0 \\
0 \\
0
\end{bmatrix}.
\end{align}
In particular, the random key $ K_2 $ will be treated as a fixed linear combination of the secret messages and the true random key $ K_1 $, to maintain total number $ N-\sum_{i\in[K]}R_i $ of  random keys in this design. The coefficients $ \alpha_{i,j} $ are used to make the system of linear equations solvable, whose physical meaning is a relaxation of the reuse of the same share, as it can be multiplied with different coefficients for different secret messages. So in order to carry on the encoding for the shares, the above linear equations can be transformed into the following:
\begin{align}\label{system_linear_equation}
	\begin{bmatrix}
		0 & \alpha_{1,1} & 0 & 0 & 0 \\
		0 & 0 & \alpha_{1,2} & 0 & 0 \\
		0 & 0 & 0 & \alpha_{1,3} & 0 \\
		1 & 0 & 0 & \alpha_{2,3} & 0 \\
		2 & 0 & 0 & 0 & \alpha_{2,4}
	\end{bmatrix}\cdot\begin{bmatrix}
	K_2 \\
	Y_1 \\
	Y_2 \\
	Y_3 \\
	Y_4
\end{bmatrix}=-\begin{bmatrix}
1 & 1 & 1 & 0 \\
1 & 2 & 2^2 & 0 \\
1 & 3 & 3^2 & 0 \\
0 & 0 & 0 & 1 \\
0 & 0 & 0 & 1
\end{bmatrix}\cdot\begin{bmatrix}
W_{1,1}\\
W_{1,2}\\
K_1\\
W_2
\end{bmatrix}
\end{align}
The feasibility is based on the full rank property of the LHS matrix, which is then by an existence lemma \cite[Lemma 1]{Khalesi2021TheCR} for the coefficients $ \alpha_{i,j} $. The decoding function relies on the individual decoding function of the initial Shamir's scheme. 

Such idea to reuse the same share for multiple secret messages simultaneously is firstly applied in \cite[Section IV]{Soleymani2021DistributedMS}, where the authors want significant overlaps of the share with some users. As they consider some specific access structures, the coefficient $ \alpha_{i,j} $ is fixed to be one and the resulting system of linear equations is still solvable, i.e., the LHS matrix in \eqref{system_linear_equation} is full rank. In \cite{Khalesi2021TheCR}, the authors consider arbitrary access structures, and it turns out that as long as the rate tuple satisfies the capacity region \eqref{weak_capacity_1} and \eqref{weak_capacity_2}, and the field size is bigger than $ \max_{i\in[K]}|A_i| $, there exist such coefficients to make the encoding procedure feasible.

\subsection{Comparison}
To begin with, recall that when a DMUSS scheme satisfies both the correction condition \eqref{def_correctness_condition} and the perfect privacy condition \eqref{def_perfect_condition}, the secret messages have to be mutually independent. When replaced with the weak privacy condition \eqref{weak_privacy}, however, the secret messages must be pairwise independent.

The reason is that, for example, if the correctness condition $ H(W_1|Y_{A_1})=0 $ and the weak privacy condition $ H(W_{2}|Y_{A_1})=H(W_{2}) $ hold, then we have
\begin{align}
	H(W_{2}|W_1)=H(W_{2}).
\end{align}
There are total $ K(K-1) $ such formulas, and they won't deduce anything further. Then when we assume that secret messages are mutually independent in \eqref{def_independent}, it is actually a stronger assumption.

Then consider the achievability proof of \cite{Khalesi2021TheCR} and our proposed algorithm in Section V.B.. We make some statements as follows:
\begin{enumerate}
	\item \cite{Khalesi2021TheCR} designs the encoding function based on individual Shamir's schemes, forming a system of linear equations to reuse the same share for multiple secret messages simultaneously. While we design the decoding function using different indeterminates based on the fact that the decoding collection of each secret message has only one set, somehow generalize the Shamir's scheme. In this way, the perfect privacy condition is equivalent to the full rank property of some matrices in \eqref{ek_prime}.
	\item The method in \cite{Khalesi2021TheCR} implies that all shares are mutually independent, as from the decoding procedure \eqref{initial_system_linear} all $ N $ shares can recover all $ K $ secret messages and $ N-\sum_{i\in[K]}R_i $ true random keys, thus make the weak privacy condition hold. While our algorithm builds from the assumption that all shares are mutually independent. In this sense, the decoding function of \cite{Khalesi2021TheCR} also fits in our algorithm.
	\item The decoding function of \cite{Khalesi2021TheCR} relies on the decoding function of the original Shamir's scheme. The encoding function of ours depends on the Gaussian-Elimination towards the generator matrix for decoding.
	\item \cite{Khalesi2021TheCR} needs an existing lemma to justify the encoding procedure, similarly, our method needs Hall's marriage theorem to show the existence of row index sets $ C_i^k $ and an existing lemma to prove the solution to the decoding function.
\end{enumerate}

 \section{Conclusions}\label{conclusion}
We have found the capacity region of the DMUSS problem under the perfect privacy condition. We use a pattern from single secret sharing regime to finish the converse part, saying that the size of the secret is upper bounded by the sum of sizes of non-colluding shares. Then any integral rate tuple satisfying the converse corresponds to a DMUSS scheme by our proposed algorithm, whose core idea is to design the decoding function. As a result, the achievable rate tuple and its converse meet, yielding the capacity region.

\appendices

%\section{proofs of the main results}
%\subsection{proof of Lemma \ref{lemma_1}}\label{proof_lemma_1}

\bibliographystyle{unsrt}
\bibliography{ref}
\end{document}